\documentclass[aps,prb,twocolumn,groupedaddress,floatfix,showpacs,preprintnumbers,amsmath,amssymb]{revtex4}

\usepackage{graphicx,epsfig}% Include figure files
\begin{document}

\title{
Dynamical spin chirality and spin anisotropy in gapped S=1/2 quantum systems
}
\author{J.E. Lorenzo}
\affiliation{Laboratoire de Cristallographie, CNRS, BP 166, 38042 Grenoble C\'edex 09, France}
\author{C. Boullier, L.P. Regnault}
\affiliation{Commissariat \`{a} l'Energie Atomique,
D\'{e}partement de Recherche Fondamentale sur la Mati\`{e}re Condens\'{e}e,
SPSMS/MDN, 38054 Grenoble cedex 9, France}
\author{U. Ammerahl}
\affiliation{2. Physikalisches Institut, Universit\"at zu K\"oln, D-50937 K\"oln, Germany}
\author{A. Revcolevschi}
\affiliation{Laboratoire de Physico-Chimie des Solides, Universit\'e Paris-Sud, 91405 Orsay C\'edex, France}
\date{\today}

\begin{abstract}

Low dimensionality quantum spin systems constitute an ideal built-in laboratory to study fundamental aspects of solid state physics. By engineering suitable compounds, fundamental theories have been tested during the past decades and many studies are still underway. Quantum phase transitions, possible coupling mechanisms to explain high-T$_C$ superconductivity, ring exchange and orbital and spin currents, Luttinger liquids and Bose Einstein condensation are among the matters studied in this fascinating land of quantum systems. Here we add two new values to this extensive list, that are the study of the spin anisotropy in spin-singlet ground state compounds and the study of magnetic chirality, as measured by inelastic polarized neutron scattering techniques. To this end we have used a paramagnetic spin singlet ground state compound and discussed in detail the scattering properties of the first excited state, a spin triplet. In-plane and out of plane magnetic fluctuations are measured to be anisotropic and further discussed in the light of the current hypothesis of spin-orbit coupling. We show that under appropriate conditions of magnetic field and neutron polarization, the \textit{trivial} magnetic chirality selects only one of the Zeeman splitted triplet states for scattering and erases the other one that posses opposite helicity. Our analysis pertains to previous studies on dynamical magnetic chirality and chiral critical exponents, where the ground state is chiral itself, the so-called \textit{non-trivial} dynamical magnetic chirality. As it turns out, both \textit{trivial} and \textit{non-trivial} dynamical magnetic chirality have identical selection rules for inelastic polarized neutron scattering experiments and it is not at all evident that they can be distinguished in a paramagnetic compound. 

\end{abstract}


\maketitle

\section{Introduction}

Spiral or helix-type arrangements are well known geometrical examples of chirality, of crucial importance in life, the relevant feature of these geometries being the \textit{handedness} (left or right) of the arrangement\cite{Barron86}. Trivial symmetry considerations dictate that structural chirality is time-reversal even and parity-reversal odd, $PT=-+$ (or also called $P-$asymmetry), whereas magnetic chirality breaks the invariance of both time and parity, $PT=--$ (or $PT-$asymmetry). In general terms, chirality results from competing interactions-induce-frustration and is a property related to parity violation, whether it is at the level of atomic arrangement within a molecule or to the presence of special electronic configurations within a given atom. Magnetoelectric multipoles such as polar toroidal moment, magnetic quadrupole and polar toroidal octupole all have the same $PT-$breaking symmetry. For completeness, spin polarization, i.e., magnetic moments, has $PT=+-$ symmetry properties (or $T-$asymmetry)\cite{DiMatteo05}.

%It is important to realize that the pair correlation functions are related to the imaginary part of a \textit{general} susceptibility through the fluctuation-dissipation theorem. There is thus a substantial difference between the symmetry of the object and the symmetry properties of the correlation function, and here the distinction to be made is between $\omega=0$ (or elastic) from $\omega=\neq 0$ (or inelastic) correlations. In particular \textit{classical} magnetic correlations, if described by propagating fields (spin waves), are entities having both $PT=+-$ and $PT=--$ symmetry features. Following from the above discussion it is straightforward to realize that the observation of the latter does not necessarily imply that the original objects or fluctuations are magneto-chiral. The opposite, however, holds true.

The importance of the spin-chirality in accounting for a certain number of properties has been put forward in the context of, for instance, strongly correlated electron systems\cite{Wen89,Taguchi00,Taniguchi04}. In the doped planar cuprates, chiral spin fluctuations have been speculated to play a central role in establishing the normal-state properties. Spin chiral fluctuations in insulating planar cuprates have been measured by Raman scattering where it has been shown that the fluctuations of the spin chiral operator $ \bf{S}_i \cdot (\bf{S}_j \times \bf{S}_k)$ contribute in the A$_2$ scattering geometry \cite{Shastry90,Sulewski91}.  In order to explain the anomalies in the pseudogap phase and eventually shed light into the superconducting order parameter of the high-T$_C$ cuprates, a number of propositions have appeared in the literature. The first one\cite{Varma97} involves orbital currents of the form $d_{x^2} - d_{y^2} +ix$ ($x=s,d_{xy}$ or $p_x \pm py$ yielding a k=0 $T-$odd component). This theory has received the support of some experimental claims: (a) dichroic studies of the photoemission signal\cite{Kaminski02} and (b) in neutron polarization studies \cite{Fauque06}. Finally, Chakravarty \textit{et al.}\cite{Chakravarty01} have proposed a $d_{x^2} - d_{y^2}$ density wave state as order parameter that breaks $P-$, $T-$reversal symmetries, as well as translation by one lattice spacing and rotation by $\pi/2$. This idea has equally received some support from polarized neutron scattering experiments\cite{Mook02,Mook04} although Stock \textit{et al.}\cite{Stock02} found no indication of this phase in the elastic channel using nonpolarized neutron beams suggesting that its contribution might be inelastic. \\

In much the same way as for the symmetry of the operators, one can define the symmetry properties of the pair correlation functions, as measured in a polarized neutron scattering experiment. Following Gukasov \cite{Gukasov99}, a further distinction between  $\omega=0$ (or elastic) from $\omega \neq 0$ (or inelastic) correlations ought to be made. $P-$asymmetry is an intrinsic property of non-centrosymmetric crystals and yields antisymmetric correlations in the (quasi-)elastic channel, alone. An external magnetic field can favor magnetic domains of the appropriate direction and hence induce macroscopic $T-$asymmetry in a ferromagnet \cite{Note1}. Conversely, the paramagnetic ground state does not violate $T-$symmetry unless a magnetic field that induces a homogeneous magnetization is applied. This magnetic field-induced ground states do not violate $P-$symmetry in the elastic channel and $T-$asymmetry develops in the inelastic scattering channel alone. These magnetic inelastic excitations\cite{Note10} are described as spin precessions around the quantization axis or in the case of paramagnets, the magnetic field. The direction of precession is given by the standard algebra of spinors \cite{Note11}.  Thus, magnetic excitations have a built-in $P-$violation regardless of whether or not the structure violates $P-$symmetry. This is what we shall call \textit{trivial}\cite{Note12} dynamic magneto-chirality.

%This concept is key to understand the results of this paper and perhaps of other published papers.

The question of how magnetic chirality appears in the neutron scattering experiments has been theoretically tackled since the aurora of this technique\cite{Blume63,Maleyev63}. The elastic case, spiral magnetic arrangements, have been widely studied in polarized neutron scattering experiments and the formalism can be checked in neutron scattering textbooks \cite{Squires78,Lovesey84a}. The possibility of observation of \textit{nontrivial} magnetic order by neutron scattering experiments, tensor-like multipolar orderings, was first formulated by Barzykin and Gor'kov \cite{Barzykin93} and later developed by Maleyev \cite{Maleyev95,Maleyev00,Maleyev02,Maleyev04} and applied to chiral compounds where the chirality clearly arises from magnetic frustration. 

\textit{Trivial}\cite{Note12} dynamic magneto-chirality has been observed in the magnetic excitations of ferromagnetic compounds. Very recently experimental evidence for chirality in the 1D S=1/2 quantum Ising antiferromagnet CsCoBr$_3$ has been detected\cite{Braun05}. Excitations corresponds to the flipping of a single spin, thus creating a domain wall, and the propagation of two solitons in both directions of the chain. As this soliton can be placed anywhere along the chain, the resulting state is highly degenerate. The propagation involves the coherent rotation of spins at next-nearest neighbor sites, the handedness of the rotation being opposite for the two solitons. An infinitely small external magnetic field is going to remove the degeneracy between both states, thus allowing the observation of a net chirality in the polarized neutron scattering experiments. This type of dynamic magneto-chirality has predicted by the theory\cite{Braun96} and boldly deduced in unpolarized neutron scattering experiments under magnetic field\cite{Asano02} is called \textit{hidden} by the authors\cite{Braun05}.\\ 

Recently the quest for magnetic fluctuations issued from a chiral spin arrangement has raised a lot of interest. Following Kawamura's conjecture \cite{Kawamura98}, the magnetic phase transitions of \textit{chiral} magnetic compounds should belong to a new universality classes (the chirality universality class), with its own order parameters and novel critical exponents. Plakhty's group have conducted inelastic neutron scattering experiments in some well known chiral compounds (in the triangular lattice antiferromagnets CsMnBr$_3$ and CsNiCl$_3$ \cite{Maleyev98,Plakhty99,Plakhty01} and in the helimagnetic phase of Ho \cite{Plakhty01}). Energy scans of the quasi-elastic scattering were carried out right above the phase transition temperature and under a polarizing field of 1-3T parallel to the momentum transfer wave vector, ${\bf H} \parallel {\bf Q}$. From the difference between the neutron counts for $\uparrow$ and $\downarrow$ neutron channels they claimed to have shown the presence of \textit{dynamic} spin chirality and associated critical exponents above T$_N$ (in the paramagnetic phase). In a parallel work, Roessli and coworkers \cite{Roessli02} have shown the presence of \textit{dynamic} spin chirality in MnSi. MnSi is a single-handed spiral ferromagnet, with a very small modulation wave vector \textbf{q}=($\epsilon, \epsilon, \epsilon$), with $\epsilon$=0.017. In this case the experiment consisted in q-scans at fixed energy and no polarizing magnetic field (except for a small guiding field of 10 Oe). The critical exponent $\nu \approx$ 0.67 \cite{Roessli02,Grigoriev05} is rather close to the value expected for chiral symmetry\cite{Kawamura98} and the results of polarized neutron small angle scattering have shown that the diffuse scattering looks like half moons oriented along the incident neutron polarization\cite{Grigoriev05}. In both cases, the spin chirality was extracted by performing the difference between the intensity collected at polarizing $\uparrow$ and $\downarrow$ incident neutron beams at non-null energy transferred. We define \textit{non-trivial} magnetic dynamical chirality as that $PT=--$ component of the excitations arising from an antiymmetric vector arrangement, ${\bf C=S_i \times S_j}$ or from electron spin currents that may be present in the compound.

Both \textit{trivial} and \textit{non-trivial} magnetic dynamical chirality neutron scattering cross-sections share the same selection rules, the former being more stringent than the latter. From the experiments carried out in these systems it is far from obvious how one can actually discriminate both contributions. This may turn out to be difficult as most of these measurements (except for those on MnSi) rely upon the application of a rather strong magnetic field (3-4T) in the close neighborhood of T$_N$. And it is in this region where a small perturbing field is going to have the largest effect. Therefore, the question of how to separate the sought \textit{non-trivial} magnetic dynamical chirality from the \textit{trivial} part should be addressed prior to any further claim on the criticality of the chiral fluctuations.

In this paper we address the issue of the the measurement of \textit{trivial} dynamic magneto-chirality. The choice of the compound, Sr$_{14}$Cu$_{24}$O$_{41}$, is not purposeless: it represents a suitable example of a two interpenetrating, non-interacting spin-liquids compound where quantum spin fluctuations are seen to survive up to room temperature. We thereby present a detailed polarized inelastic neutron scattering study of the excitations in the paramagnetic compound Sr$_{14}$Cu$_{24}$O$_{41}$. This family of compounds exhibit a composite structure made up of a sub-lattice of S=1/2 spin chains and a sublattice of S=1/2 spin-ladders. The magnetic excitations of the chain sublattice will be addressed here, with a special emphasis in \textit{(a)} the anisotropy of the spin excitations of a degenerate spin triplet and \textit{(b)} a thorough study of the spin-spin antisymmetric correlation functions. In view of the close relation between the observed anisotropy of the magnetic excitations and the occurrence of a\textit{non-trivial} magnetic chirality we have decided to study both in this paper. This paper is structured as follows: We first discuss the experimental details and the basic features of longitudinal neutron polarization analysis needed to follow the results of this paper. Next, the anisotropy of the magnetic excitations is characterized in two different experiments \textit{(i)} by measuring the intensity of the spin-triplet components under magnetic field and \textit{(ii)} by performing a neutron polarization analysis and extracting the in and out-of the scattering plane spin-spin correlation functions. Polarization studies under magnetic field of the spin-excitations, $H \parallel \bf{Q}$, allow to measure the influence of the antisymmetric spin-spin correlation (or \textit{trivial chirality}) onto the scattering cross section. Finally we conclude by hinting on the possible origin of the anisotropy of the spin-triplet correlation functions and on the impact of our studies on the observation of possible dynamic chirality features (or \textit{non-trivial chirality}) in a neutron scattering experiment.

\section{Experimental details}

\subsection{Longitudinal Polarization Analysis}

Longitudinal Polarization Analysis (LPA) \cite{Moon69} has been largely used to study magnetic excitations in condensed matter. It consists of creating a spin-polarized incident neutron beam (the polarization rate of the incident beam is {\bf P$_0$}) along a given direction and measuring the number of neutrons scattered in the same direction and in the different polarization states, parallel or antiparallel with respect to the incident neutron polarization settings. If each of the polarization states is labelled by the direction of the neutron spin, $+$ (or $\uparrow$) and $-$ (or $\downarrow$), then the different cross sections are denoted by the pairs $(++)$ and $(--)$ for the \emph{non-spin-flip} (NSF) processes and $(+-)$ and $(-+)$ for the \emph{spin-flip} (SF) ones. It can be easily shown that the most suitable reference system for the neutron polarization analysis refers to the scattering vector, {\bf Q}, and thus we define the components as: $x \parallel {\bf Q}$, $y \perp {\bf Q}$ and $z$ vertical to the scattering plane. This method allows the determination of both nuclear and magnetic contributions by measuring the polarization cross-sections for the three different directions of {\bf P$_0$}. Theoretical equations describing cross-sections and final polarization state have been independently derived by Blume and Maleyev~\cite{Blume63,Maleyev63}. These equations and, in general, the LPA methodology has been recently revisited by us \cite{Regnault05}, and in what follows only the final equations will be given

\begin{eqnarray}
\sigma_x^{\pm \pm} & \propto &  N \nonumber \\
\sigma_x^{\pm \mp} & \propto &   M_{y} + M_{z} \mp M_{ch} \nonumber \\
\sigma_y^{\pm \pm} & \propto &  N + M_{y} \pm R_y \nonumber\\
\sigma_y^{\pm \mp} & \propto &  M_{z} \nonumber \\
\sigma_z^{\pm \pm} & \propto &  N + M_{z} \pm R_z \nonumber \\
\sigma_z^{\pm \mp} & \propto &    M_{y}
\end{eqnarray}
where $\sigma_\alpha^{\beta,\gamma}$ is the short form of $(d^2\sigma/d\Omega d\omega)^{\beta,\gamma}(P_0 \parallel \alpha)$. For completeness the unpolarized neutron scattering cross-section is $\sigma = N + M_y + M_z$. The notation of the above equations is as follows:

\begin{eqnarray}
N &=& \langle N_QN_Q^\dagger \rangle_\omega \nonumber \\
M_y &=& \langle M_{Qy}M_{Qy}^\dagger \rangle_\omega \nonumber \\ 
M_z &=& \langle M_{Qz}M_{Qz}^\dagger \rangle_{\omega} \nonumber \\
M_{ch} &=& i (\langle M_{Qy}M_{Qz}^\dagger \rangle_{\omega} - \langle M_{Qz}M_{Qy}^\dagger \rangle_{\omega}) \nonumber \\
R_y &=& \langle N_{Q}M_{Qy}^\dagger \rangle_{\omega} + \langle M_{Qy}N_{Q}^\dagger \rangle_{\omega} \nonumber \\
R_z &=& \langle N_{Q}M_{Qz}^\dagger \rangle_{\omega} + \langle M_{Qz}N_{Q}^\dagger \rangle_{\omega} 
\end{eqnarray}
where $\langle N_QN_Q^{\dagger} \rangle_{\omega}$ and $\langle M_{Q\alpha}M_{Q\alpha}^\dagger \rangle_{\omega}$ ($\alpha$ = y, z) are the space and time Fourier transform of the nuclear-nuclear and spin-spin correlation functions, respectively. $R_y$ and $R_z$ are the symmetric part of the nuclear-magnetic interference terms and $M_{ch}$ is the chiral (or antisymmetric) correlation function. It is worth noting that antisymmetric part of the interference terms and the symmetric counterpart of the chiral correlation function are not accessible by the LPA technique as the polarization of the incident neutrons results rotated after scattering by these terms. In order to access these correlation functions the use of spherical neutron polarimetry based on, e.g., "Cryopad" devices \cite{Brown93,Regnault05} is mandatory. 

Before closing this section on the LPA technique it is worth recalling that the last three correlation functions ($M_{ch}$, $R_y$ and $R_z$), if non null, generate a polarized beam, when the incoming beam in unpolarized, $P_0=0$. The resulting polarization of the scattered neutrons is along $x$, $y$ and $z$ directions, respectively, and the cross-sections for this case are 

\begin{eqnarray}
\sigma_x^{0\pm} &\propto&    N+M_{y} + M_{z} \pm M_{ch} \nonumber \\
\sigma_y^{0\pm} &\propto&    N+M_{y} + M_{z} \pm R_{y} \nonumber \\
\sigma_z^{0\pm} &\propto&    N+M_{y} + M_{z} \pm R_{z}
\end{eqnarray}
A variant of this last configuration has been used by authors in refs. \cite{Maleyev98,Plakhty99,Plakhty01} to select the chiral correlation function from the rest of symmetric contributions to the scattering cross section. The alternative experimental situation consists of producing a polarized incoming beam and no polarizarion analysis is carried out in the scattered beam.  Note that because of the symmetry of the equations, the very same terms can generate an unpolarized scattered beam out of polarized incident. $M_{ch} \equiv 1/2(\sigma_x^{-0}-\sigma_x^{+0})$ $\equiv 1/2(\sigma_x^{0-}-\sigma_x^{0+})$. Finally, one has to keep in mind that the development of the neutron scattering cross sections is completely general and exclusively based on the properties of the magnetic interaction vector and the neutron spin polarization. These equations are independent of the choice of a particular magnetic interaction (as for instance the Dzyaloshinskii-Moriya antisymmetric spin-exchange) or spin model.

\begin{figure}[!ht]
%\color{blue}
\begin{center}
\epsfig{file =  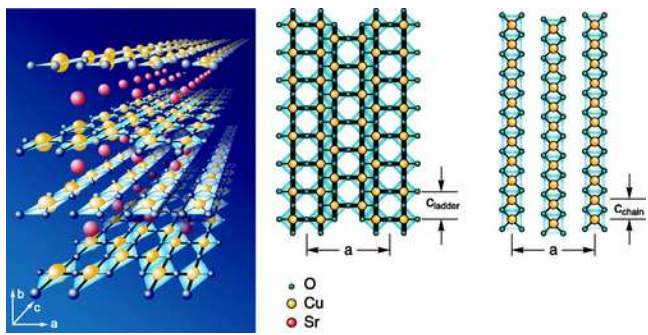, width=8.5cm }
\caption{\label{Structure} (color online) (Left) Structure of Sr$_{14}$Cu$_{24}$O$_{41}$. It is a stacking of layers of chains and layers of ladders separated by Sr (or Ca, La,  $\cdots$.). (Right) Detail of the structure of the chain sublattice and the ladder sublattice }
\end{center}
\end{figure}

\subsection{Sample description and experimental conditions}

Polarized neutron experiments were performed on the paramagnetic, spin-singlet ground state compound Sr$_{14}$Cu$_{24}$O$_{41}$. This compound displays a composite structure made up of the stacking of two distinct low dimensional Cu-O arrangements exhibiting a spin-singlet ground state\cite{VanSmaalen99}. The first sub-system is a 1-dimensional lattice of edge sharing CuO$_2$-chains and the second one is a 2D system of two-leg ladders, Cu$_2$O$_3$, the stacking direction being the $b-$axis. Lattice parameters for the chain sublattice are $a$=11.53 $\dot{A}$, $b$=13.37 $\dot{A}$, $c_c$=3.93 $\dot{A}$. The admixture of both sub-systems originates a superstructure with a nearly commensurate ratio of chain and ladder units along the $c-$direction, $10c_l \approx 7c_c$ that results in a rather large lattice parameter $c=27.52$ \AA~for the supercell. A representation of both atomic positions and magnetic system of chains and ladders is given on Figure~\ref{Structure}. As first reported by McCarron III \textit{et al.} \cite{McCarron88} and latter refined within the superspace formalism by Frost-Jensen \textit{et al.} \cite{FrostJensen97b}, the CuO$_2$ sublattice is described in the $Amma$ space group while the SrCu$_2$O$_3$ sublattice can be described in space group $Fmmm$. In-between these two types of copper oxide layers, the layers of Sr atoms are interleaved (Fig. l c) and a number of dopings have been studied \cite{VanSmaalen99}. $A_{14}$Cu$_{24}$O$_{41}$ ($A$ = Sr, Ca, La, $\cdots$) is the only known spin ladder material supporting carriers doping. Interestingly, by substituting Sr$^{2+}$ by La$^{3+}$ the number of holes in the unit greatly diminishes whereas Ca$^{2+}$ doping favors a transit of holes from the chain sublattice to ladder sublattice. Note that on going from pure Sr to pure Ca the $b-$ lattice parameter shrinks by 1 \AA, and for $x \approx 13.6$ it has been shown to exhibit superconductivity under pressure (3-5 kbar)\cite{Uehara96,Isobe98}. 

\begin{figure}[!ht]
%\color{blue}
\begin{center}
\epsfig{file = 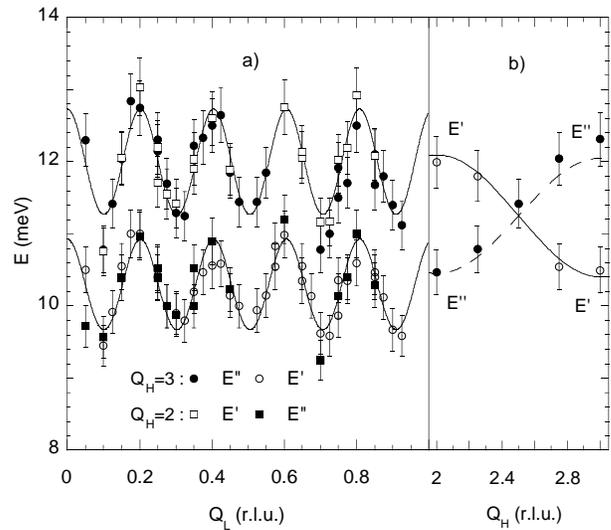, width=8cm }
\caption{\label{Dispersion} %\textbf{Chain-Dispersion.eps}. 
Dispersion of the chain excitations along the $a-$ and $c-$directions. The continuous line is the result of a fit based on a isolated dimers model \cite{Regnault99}. Experiments reported here have been carried out at \textbf{Q}=(-2.5, 0, 0.25), where the two branches merge and become degenerated.}
\end{center}
\end{figure}

\begin{figure}[!ht]
%\color{blue}
\begin{center}
\epsfig{file =  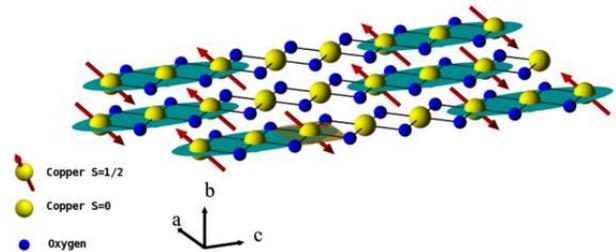, width=8.5cm }
\caption{\label{Structure_chains} %\textbf{Sr14Cu24O41-Chaines3D-tris.eps}. 
(color online) Hole ordering and the concomitant magnetic arrangement as it has been determined at low temperatures by inelastic neutron scattering experiments \cite{Regnault99}. The green shadowed area represents the spatial extension of the spin-dimer, where spin-spin (super)exchange takes place through a S=0 Cu. }
\end{center}
\end{figure}

In this paper, we have studied the chain sub-system. According to previous studies\cite{Matsuda96b,Regnault99,Matsuda99} the inelastic spectrum of chain system has been investigated and a well defined magnetic gap is observed around 11 meV for temperatures below $T<\Delta/k_B$. As there are 2 symmetry different, hardly interacting chains per unit cell along $a$, two distinct triplet states appear (Figure~\ref{Dispersion}). To deal with the difficulty of well separating the two excited states, the experiment was carried out at \textbf{Q} = (-2.5, 0, 0.25) (and symmetry related positions) where dispersion curves of the two distinct triplet states cross and a single mode appears at this position. As it can be seen in figure \ref{Structure_chains}, the magnetic ground state observed at low temperatures results from the peculiar charge ordering (hole ordering) developing in this compound where the extra holes serve to form Zhang-Rice singlets\cite{ZhangRice88,Matsuda96b} at given Cu positions. Indeed this charge ordering develops continuously \textit{without} a real symmetry-breaking phase transition and its origin still defies understanding. Remarkably, the refinement of the inelastic neutron scattering data has yielded a unique solution for the location of the magnetic moments, and from there the determination of the charge order pattern\cite{Regnault99,Matsuda99}. This is in striking contrasts with regular X-ray and neutron diffraction studies that require the refinement of the intensities of the peaks from both the ladders and the chain sublattices, as well as of the interference peaks between both substructures.  This extraordinary complexity hinders the realization of reliable crystallographic refinements\cite{Etrillard04,Braden04} and the whole issue is still under active debates\cite{Zimmermann04}. Therefore the study of the magnetic excitations, where both sublattices have different energy ranges with hardly no magnetic interference between them, offers the possibility of carrying out a sort of \textit{unconventional} crystallography of the charges themselves, a matter that is otherwise challenging. The neutron scattering data on the pure Sr$_{14}$Cu$_{24}$O$_{41}$ agrees with the picture of 6-holes per chain per unit cell and hole-empty ladder sublattice. Recently, Abbamonte \textit{et~al.} have given evidences for the presence of a hole crystallization in the ladder sublattice due to long-range Coulomb repulsion and without lattice distortion\cite{Abbamonte02,Abbamonte04}. This feature, first revealed in the pure compound through the presence of a 5$c_l$ ordering wave vector, has been further detected in the Ca-doped Sr$_{3}$Ca$_{11}$Cu$_{24}$O$_{41}$ with a 3$c_l$ modulation\cite{Rusydi05}. In view of these results it is certain that the model used to analyze the $q-$dependence of the inelastic neutron scattering data and thus to locate the holes in the of the chain sublattice\cite{Matsuda96b,Regnault99,Matsuda99} should be improved. However it does not cast any doubt that magnetic excitations are issued from spin-singlet to spin-triplet transitions and this feature will be utilize all through this paper. \\

Our Sr$_{14}$Cu$_{24}$O$_{41}$ sample was cut from an ingot grown by the traveling solvent zone method under a pressure of 3 bars of oxygen~\cite{Revco97}. The sample used in the inelastic neutron scattering experiment is made up of a set of five cylindrically shaped single crystals of volume 5$\times$5$\times$10 mm$^3$ with {\bf c} axis along the rod with a misalignment among the five single crystals of the order of $\pm$0.5 $^\circ$.

Experiments were carried out on the CRG three-axis spectrometer IN22 at the Institute Laue-Langevin set in 3 different monochromator-analyzer configurations:

\begin{itemize}
\item [1.] Pyrolitic graphite(PG) - PG, for standard unpolarized studies.
\item [2.] Heusler-Heusler, for full polarization analysis studies.
\item [3.] and PG-Heusler for experiments where only the polarization of the scattered beam is analyzed, out of a unpolarized incoming neutrons beam. 
\end{itemize}

and three different sample environment configurations,

\begin{itemize}
\item  [1.] ILL-type orange cryostat, available to cover the range 1.4-300K.
\item  [2.] Vertical superconducting magnet of 12T. For the polarized neutron scattering experiments a 6T sueprconducting magnet was used and the maximum polarizing field was of 3T.
\item  [3.] Horizontal superconducting magnet of 4T.
 \end{itemize}

\begin{figure}[!ht]
%\color{blue}
\begin{center}
\epsfig{file =  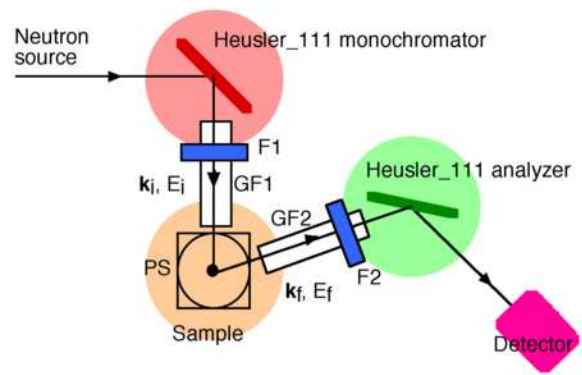, width=8.5cm}
\caption{\label{Instrument}
(color online) Sketch of the experimental device for the type-2 configuration. Neutrons were monochromatized and vertically polarized by means of a Heusler crystal, say in the + state; the flipper (F1) allows for a $+ \rightarrow -$ neutron \textit{flip} if required. After being conducted and preserved from depolarization by a guide-field (GF1), neutrons are aligned along a given polarization direction with the help of, either a modified Helmholtz 4 set-coils (3 coils horizontally spanning 120 degrees each and 1 vertical) or a horizontal/vertical superconducting magnet. After interaction with the sample, neutrons follow through a guide-field (GF2) and can be vertically flipped (F2) if required ($- \rightarrow +$). Finally a second Heusler crystal, having the same setting as the first one, selects the corresponding polarization channel and energy analyzed the neutrons.}
\end{center}
\end{figure}

Fig. \ref{Instrument} shows a sketch of the experimental device. A particular care have been taken to determine the flipping ratio for the different field configurations. In order to minimize the effect of the variation of the cryomagnet strayfields during the course of the scans, the flipper currents were tuned to operate at 3$\pi$ flipping, instead of the most classical $\pi$ flipping. This operation mode has demanded the development of a special water-cooled flipper as the current for the 3$\pi$-flipping is three times larger than that of the $\pi$ flipping. Ideally one would like to run the flipper at higher flipping angles. However this implies a con-commitant increase of the current in the flippers that gives rise to an augmentation of the thermal charge, difficult to dissipate in such compact devices. In the 3$\pi$-flipping mode, a flipping ratio as high as 30 was acheived at 2.662 \AA$^{-1}$.

Inelastic scans were performed at fixed final wave vector k$_f$=2.662 or 3.84 $\dot{A}^{-1}$ and a 40-mm-thick graphite filter was used after the sample to minimize higher-order flux contamination. The neutron measurements were performed with the ({\bf a},{\bf c}) crystallographic plane parallel to the scattering plane. The magnetic field was applied either vertical to the scattering plane (parallel to the $b-$axis) or parallel to the scattering vector. Only the chain-sublattice magnetic excitations are here reported and the experiments were performed in the energy transfer range 8-15 meV, with typical resolution of the order of 1.5 meV (FWHM). The ladder excitations appear to have a spin-single to spin-triplet gap of 31meV and an analogous study will be reported elsewhere \cite{Boullier06}.

\subsection{Description of the spin triplet and correlation functions}

The spin pairing takes place through a non-magnetic Cu (Zhang-Rice singlet), the description of the spin excitations in terms of the lowest lying energy levels suffices to account for the inelastic neutron scattering data (below 40meV). Indeed a full description of the magnetic dimer unit involves the electronic orbitals of not less than 3 CuO$_2$ units (see figure \ref{Structure_chains}) which, in view of the large degrees of freedom involved, may result in an important number of excited states.  Despite the apparent complexity of the spin chains sublattice ground state, the theoretical framework to account for the lowest lying excited states is rather trivial. It is certain that, at some point in the analysis, a full description of the electronic states and their hybridizations ought to be invoked. This is particularly true if one wants to explain the anisotropy of the spin susceptibility \cite{Boullier05}.

At sufficiently low temperatures, T$< \Delta/k_B$, the paramagnetic excitations become very well defined and the recorded spectra are resolution limited. Spin singlet-to-triplet transitions can be evaluated by using the equation

\begin{eqnarray}
\frac{d^2\sigma}{d\Omega d\omega} & = & r_0^2 \frac{k^\prime}{k} \sum_{\alpha \beta} (\delta_{\alpha \beta} - \tilde{k}_\alpha \tilde{k}_\beta)  \nonumber \\ 
& & \sum_{v,v^\prime} p_n \langle \Gamma_n v|\hat{Q}^+_\alpha |\Gamma_n^\prime v^\prime \rangle \langle \Gamma_n^\prime v^\prime|\hat{Q}_\beta |\Gamma_n v \rangle
\end{eqnarray}
and therefore the scattering cross section is proportional to $\langle \bf{0} | \bf{\hat{S}} | \bf{1} \rangle \, \langle \bf{1} | \bf{\hat{S}} | \bf{0} \rangle $, with $| \bf{0} \rangle \equiv | 00 \rangle \equiv |\! \uparrow \downarrow \rangle - |\!\downarrow \uparrow \rangle$ the spin singlet and $| \bf{1} \rangle \equiv \{ | 11 \rangle, | 10 \rangle, | 1\bar{1} \rangle \} \equiv \{ |\! \uparrow \uparrow \rangle, |\!\uparrow \downarrow \rangle + |\!\downarrow \uparrow  \rangle, |\!\downarrow \downarrow \rangle \}$, the spin-triplet. By using the traditional axis for quantum mechanic calculations ($\sigma_X |\!\uparrow \rangle = |\!\downarrow \rangle$; $\sigma_Y |\!\uparrow \rangle = i|\!\downarrow \rangle$; $\sigma_Z |\!\uparrow \rangle=|\!\uparrow \rangle $, and identically for $|\!\downarrow \rangle$) the scattering probabilities are found to be those in Table \ref{tab:table1}.

\begin{table}
\caption{\label{tab:table1} Value of $\langle \bf{0} | \bf{\hat{S}} | \bf{1} \rangle$ for the different components of the triplet, $\bf{1}$. These are labelled, following the notation $| S s_z \rangle$. }
\begin{ruledtabular}
\begin{tabular}{l|ccc}
%\hline\hline
 & $S_X$ & $S_Y$ & $S_Z$ \\
\hline
$| 10 \rangle $ & 0 & 0 & $\sqrt{2}$\\
$| 11 \rangle $ & 1 & $-i$ & 0 \\
$| 1\bar{1} \rangle $ & 1 & $i$ & 0 \\
%\hline
%\hline
\end{tabular}
\end{ruledtabular}
\end{table}

\section{Results}

\subsection{Magnetic excitations without polarization analysis}

\begin{figure}[!ht]
%\color{blue}
\begin{center}
\epsfig{file =  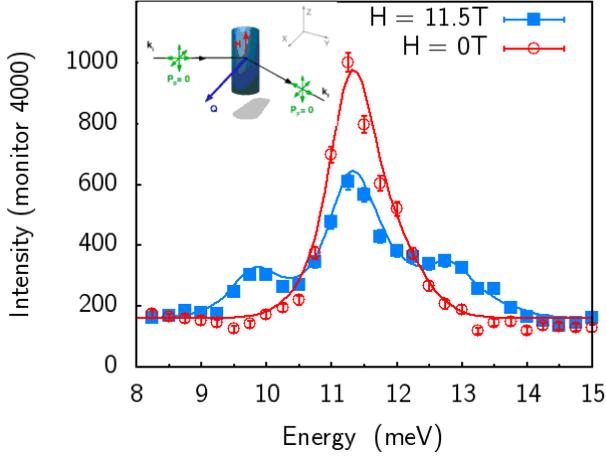, width=8.5cm }
\caption{\label{12T_P_0} %\textbf{NonPolarises-0T12TChampVertical.eps}. 
(color online) Unpolarized neutron scan of the Zeeman splitting of the triplet at \textbf{Q} = (-2.5, 0, 0.25) at H=11.5T (vertical), and comparison with H=0. The neutron polarization conditions appear as an insert. }
\end{center}
\end{figure}

\begin{figure}[!ht]
%\color{blue}
\begin{center}
\epsfig{file =  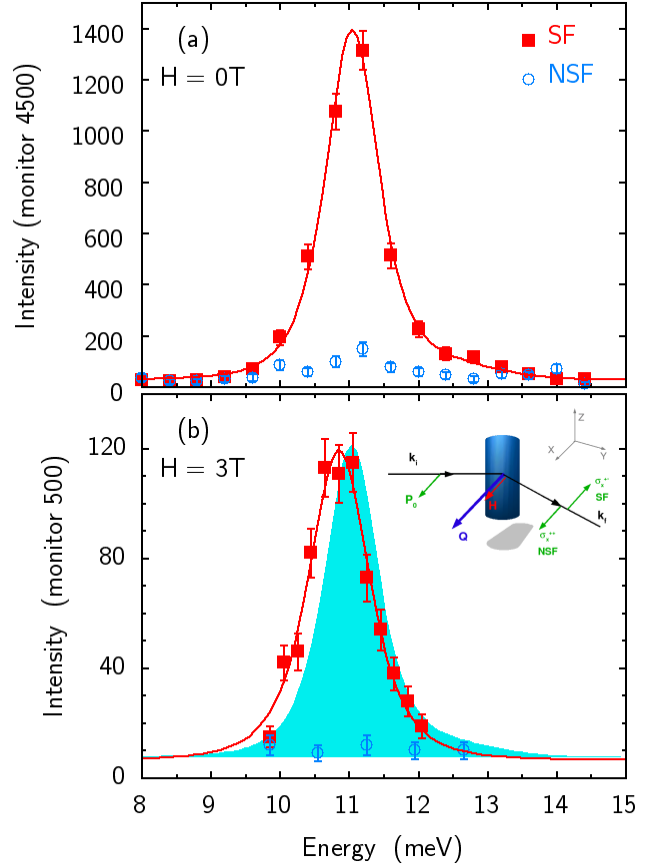, width=8.5cm}
\caption{\label{0T_H_P} %\textbf{ChainInelasticScanXX-superpose.eps}. 
(color online) (a) Polarized (P$_0 \parallel \bf{Q}$) neutrons energy scan at Q=(-2.5, 0, 0.25) and with H=0T. The scattering cross-sections are $\sigma_x^{--}$ in blue and $\sigma_x^{-+}$ in red. The fact that  $\sigma_x^{--} \approx$ 0 (except for background correction) implies that the excitations measured at this \textbf{Q}-position and energy are purely magnetic. (b) Same scan under horizontal magnetic field of 3T with $\sigma_x^{++}$ in blue and $\sigma_x^{+-}$ in red. As a comparison we have included the H=0T data as a blue shadow. Data at H=3T clearly display a shift to low energies whereas no signal appears at high energies.}
\end{center}
\end{figure}

The excitations corresponding to the spin-chain sublattice have been studied in detail\cite{Matsuda96b,Regnault99,Matsuda99} albeit without neutron polarization analysis. Under a field of 11.5T the resolution conditions are such that the three components of the Zeeman-splitted triplet can be well separated (see figure \ref{12T_P_0}). From the peak positions $\hbar\omega_1$= $\hbar\omega_0 - g_b\mu_BH_b$ = 9.8 meV, $\hbar\omega_0$= 11.3 meV and $\hbar\omega_{\bar{1}}$= $\hbar\omega_0 + g_b\mu_BH_b$ = 12.8 meV the value of the Land\'e factor perpendicular to the chain axis can be calculated, $g_{b}=$2.31$\pm$0.06, in agreement with magnetic susceptibility\cite{Matsuda96a} and ESR\cite{Matsuda96a,Kataev01a,Kataev01b} data.  No nuclear component has been detected at this \textbf{Q}-position (Figure \ref{0T_H_P}) and the scattering cross-section is proportional to $\sigma (Q,\hbar\omega_0 \pm g\mu_BH) + \sigma (Q,\hbar\omega_0)$. The direction of the magnetic field, $z$, imposes this direction to be the quantification axis which, from the table \ref{tab:table1}, implies that $Z \equiv z$. Under these circumstances and by looking at the results in Table \ref{tab:table1} one can safely conclude that $\sigma (Q,\hbar\omega_0 \pm g\mu_BH) \equiv M_y$ and $\sigma (Q,\hbar\omega_0) \equiv M_z$.\\

A careful analysis of the integrated intensities of the three peaks reveals that these are weaker than that of the degenerate H=0T integrated intensity, I$_{1,\bar{1}}$(H=11.5T)=0.35*I$_0$(H=0) for the side peaks and I$_0$(H=11.5T)=0.54*I$_0$(H=0) for the unshifted one. Therefore the ratio $M_z/(M_y(\hbar\omega_{\bar{1}})+M_y(\hbar\omega_1))$ is not 1 but rather 1.3$\pm$ 0.1, which leads to the conclusion that the magnetic fluctuations are anisotropic. These values of the intensities obtained under magnetic field, once added, yield the same amount as that obtained for the intensity of the degenerate triplet at H=0T. 

\begin{table*}
\caption{\label{tab:table2} Measurement of the magnetic susceptibility anisotropy $M_z/M_y$  of the chain sublattice at T=2.5K and H=0T}
\begin{minipage}{12cm}
\begin{tabular}{cc|ccccc|c}
\hline
$\mathbf{Q}$ & $\omega$ (meV) & $M_z$ & $M_y$ & Background & Monitor & Counting time (s) & $M_z/M_y$ \\
\hline
(-2, 0, 0.3) & 10 & 442 & 339 & 30 & 5000 & 2430 & 1.33 $\pm$ 0.11\\
           & 11.3 & 334 & 270 & 30 & 5000 & 2519 & 1.28 $\pm$ 0.15\\
(-2, 0, 0.7) & 11 & 437 & 290 & 25 & 4000 & 2010 & 1.58 $\pm$ 0.14\\
             & 11 & 127 & 101 & 38 & 4000 & 2018 & 1.29 $\pm$ 0.27\footnote{Temperature for this point was 150K} \\
(-3, 0, 0.3) & 10 & 470 & 348 & 44 & 8000 & 3998 & 1.60 $\pm$ 0.15 \\
           & 11.3 & 274 & 194 & 28 & 5000 & 2515 & 1.60 $\pm$ 0.23\\
(-3, 0, 0.7) & 10 & 598 & 477 & 50 & 5000 & 2422 & 1.22 $\pm$ 0.12\\

\hline
\end{tabular}
\end{minipage}

\vspace{2mm}

%{0.9\linewidth}
%\end{center}
\end{table*}

\subsection{Magnetic excitations under polarization analysis}

Polarized neutrons are most frequently utilized in experiments aiming at separating nuclear and magnetic contributions to the scattering. By inspecting the polarized neutron scattering equations (eqs. 1) one realizes that the configuration $P_0 \parallel \bf{Q}$, independently of the direction of the magnetic moments, is the most simple way to discriminate between both contributions. Indeed, the NSF cross-sections ($\sigma_x^{\pm \pm}$) contains the nuclear contribution alone, whereas the SF channel ($\sigma_x^{\pm \mp}$) is proportional to the components of the magnetization perpendicular to $\bf Q$, $M_{y} + M_{z} \mp M_{ch}$. Results at H=0T in figure~\ref{0T_H_P} shows that the NSF contribution is zero and therefore a pure magnetic scattering appears at the \textbf{Q}-position of the experiment. Before closing this section it is important to recall that polarization analysis implies the presence of a rather small polarizing field at the sample position (H$_p \approx$ 0.1T) in order to prevent neutron depolarization. Therefore, and strictly speaking, the sample is never at H=0T in our polarized neutron scattering experiments.

\begin{figure}
%\color{blue}
\begin{center}
\epsfig{file =  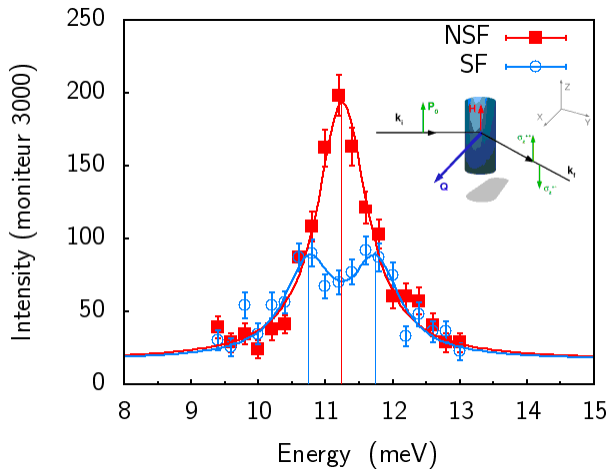, width=8.5cm}
\caption{\label{4T_V_P} (color online) Polarized inelastic spectra under a vertical magnetic field of 4T with $\sigma_z^{++}$ in red and $\sigma_z^{+-}$ in blue.}
\end{center}
\end{figure}

\subsubsection{Anisotropy in the vertical field configuration}

As already mentioned above and shown in Figure \ref{12T_P_0}, inelastic neutron scattering experiments at rather high magnetic fields have revealed a unexpected anisotropy between the $| \!1 1 \rangle$ (or $| \!1 \bar{1} \rangle$) and $|\! 1 0 \rangle$ components of the spin-triplet or, in other words, between the in-plane ($M_y$) and out-of-plane ($M_z$) spin-spin correlation functions. Identical conclusion can be drawn out from the studies at zero field and under a polarized beam. Indeed, and in the absence of a nuclear contribution (meaning that $N$ and $R_z$ are zero in equations 1) the configuration $P_0 \parallel z \perp \bf{Q}$ readily implies that the signal in the NSF channels ($\sigma_z^{\pm \pm}$) is proportional to $M_z$ whereas $M_y$ correlation functions appear in the the SF channels ($\sigma_z^{\pm \mp}$). Results are shown in Figure \ref{4T_V_P} for $\bf{Q}$=(-2.5 0 0.25). Note that the use of a magnetic field in this experiment is exclusively justified in terms of \textit{cosmetic} reasons and it does not bring relevant information other than to separate the $s_z$=1 from the $s_z = \bar{1}$ components of the triplet. Identical studies can be carried out at different \textbf{Q}-positions and data are displayed in Table \ref{tab:table2}. Regardless the \textbf{Q}-position, an anisotropy in the susceptibility of the order of 30\%, first evidenced in susceptibility measurements, is thus confirmed by our analysis of the spin-spin correlation functions. The fact that this ratio is roughly independent on Q$_c$ implies that magnetic fluctations within the $(a,c)$ plane are isotropic. The origin and magnitude of this anisotropy is puzzling and difficult to justify in terms of the current, although simplified, models of spin-spin interactions. In the case of a S=1/2 system, single ion anisotropy is zero. Other well known sources of anisotropy is the Dzyalozhinskii-Moriya (DM) antisymmetric interactions. However this antisymmetric interaction will in turn break the spin-triplet degeneracy, and such a splitting has not been observed in our experiments. Moreover DM is not allowed by symmetry in this compound. We shall come back to this point below.

\begin{figure}
%\color{blue}
\begin{center}
\epsfig{file =  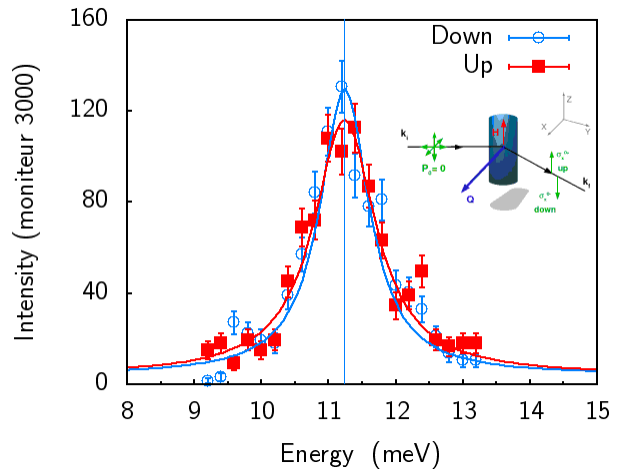, width=8.5cm}
\caption{\label{4T_V_P_0} (color online)Inelastic spectra at H=4T and with unpolarized incoming beam, $P_0=0$ (Graphite-Heusler). Cross-sections are $\sigma_z^{0+}$ in blue and $\sigma_z^{0-}$ in red.}
\end{center}
\end{figure}

\subsubsection{Study of the magneto-chiral correlation function}

Magnetic excitations are defined as spin-precessions around the quantization axis. In the absence of a magnetic field, energy minimization considerations dictate that there are equal number of spins pointing up ($n_\uparrow$) and pointing down ($n_\downarrow$)\cite{Note2}, and therefore there is not net macroscopic helicity in the system. In a paramagnet the quantization axes is defined by the magnetic field itself which, in addition, defines the beam polarization direction. In this configuration $\bf{H} \parallel \bf{Q} \parallel x$, the spin correlations that one have access to are $M_y$ and $M_z$ (symmetric) and $M_{ch}$ (antisymmetric) (see equations 1 and 2). We shall call this latter term \textit{trivial} dynamical magnetochirality (proportional to the difference $(n_\uparrow - n_\downarrow)$) in order to distinguish it from the proposal of magneto-chiral fluctuations issued from a odd-parity magnetic arrangement \cite{Maleyev98,Plakhty99,Plakhty01}. The necessary condition for its observation requires either a completely polarized neutron beam or at least one of the components of the polarization (incoming or outgoing beam) be well defined. As it is pointed out in the introduction, this condition, however, does not suffice to observe it as time reversal symmetry should be broken (macroscopically) in order to have only one of the helicities in the groundstate or at least to unbalance them. Dynamical magnetochirality should be observed in uniaxial ferro and ferrimagnets and paramagnets under an external magnetic field, either to create a single domain (in the former) or to privilege a given direction (in the latter). \\

So far we have discussed the paramagnetic state of compounds that are going to magnetically order as the temperature is decreased. A different class of paramagnetic compounds are those exhibiting a non-magnetic spin-singlet ground state down to the lowest temperatures. Here magnetic fluctuations arise from the population of the first (and beyond) excited state, a S=1 spin triplet. Apart from the compound under consideration, 1D spin chains, CuGeO$_3$\cite{Hase93}, NaV$_2$O$_5$\cite{Isobe96}, the high T$_C$ superconductors, the frustrated dimer arrangement SrCu$_2$(BO$_3$)$_2$\cite{Kageyama00} or the spin-ladder compounds, CuHpCl\cite{Hammar96,Chaboussant97}, (VO)$_2$P$_2$O$_7$\cite{Johnston87,Barnes94}, \textit{etc.}, are specially important. Paramagnetic excitations appear at the spin-triplet gap and above, and are clearly inelastic at sufficiently low temperatures.

\begin{figure}[!ht]
\begin{center}
\epsfig{file =  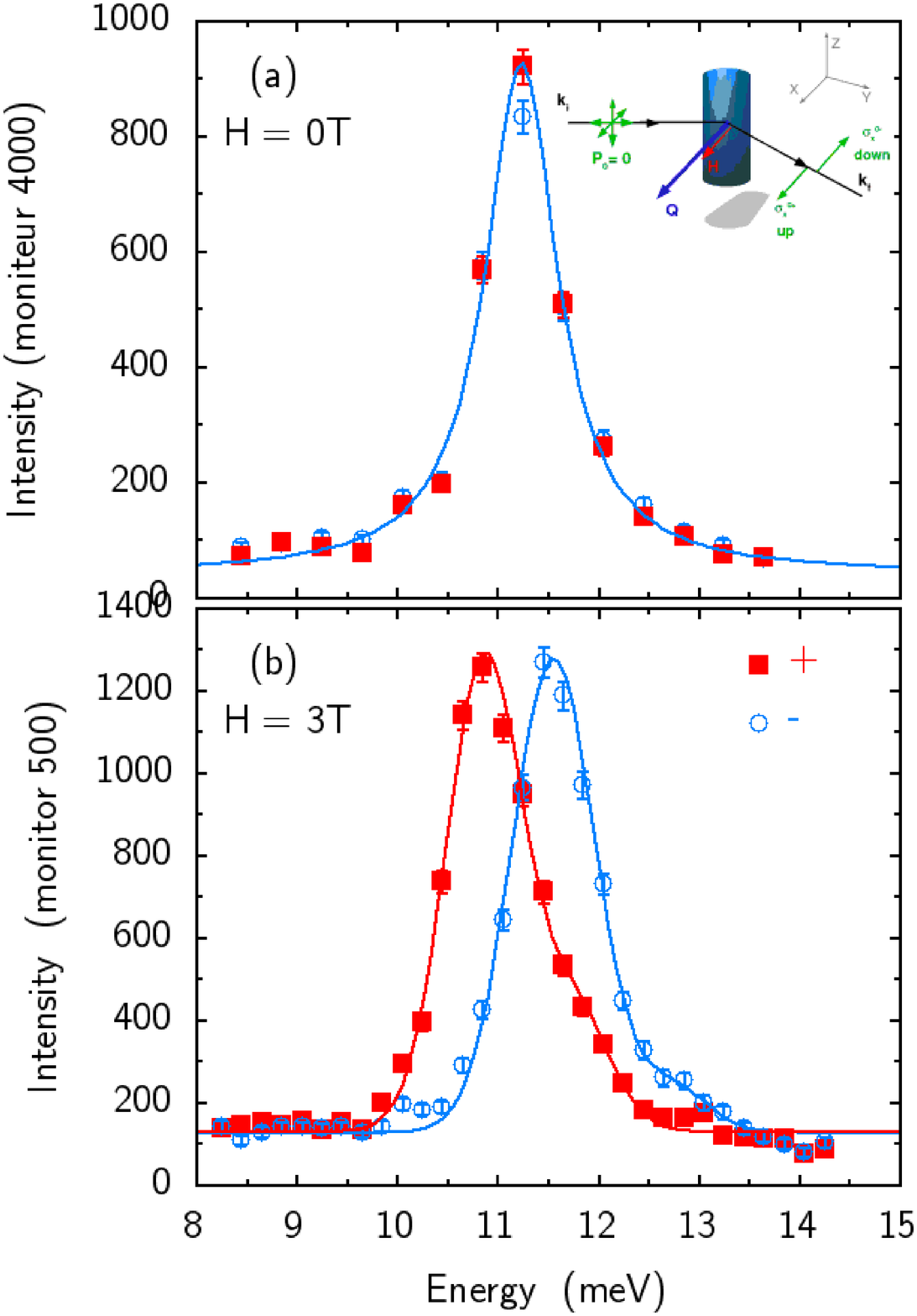, width=8.5cm}
\caption{\label{3T_H_P_0} (color online) (a)  Inelastic spectra at H=0 and with unpolarized incoming beam, $P_0=0$ (Graphite-Heusler). Cross-sections are $\sigma_x^{0+}$ in blue and $\sigma_x^{0-}$ in red. In this case the $+$ and $-$ are related to the sign of \textbf{Q}. (b) \textit{Idem} with a horizontal magnetic field H=3T oriented in the same direction that \textbf{Q}. In the presence of an applied magnetic field $\sigma_x^{0+}$ measures $|1 1 \rangle $ component of the triplet whereas $\sigma_x^{0-}$ measures $|1 \bar{1} \rangle $ component. Because of the selection rules for $P \parallel \bf{Q} \parallel M_Z$, $|1 0 \rangle $ does not appear in the spectra. }
\end{center}
\end{figure}

The spin-spin correlation functions for the spin-singlet to spin-triplet transition can be easily calculated (see table \ref{tab:table1}). Here $|1 1 \rangle$ and $|1 \bar{1} \rangle$ excitations describe right and left handed helices, respectively. The experiment was carried out with a magnetic field of 3T parallel to \textbf{Q} that imposes the quantization direction for the excitations and thus removes the degeneracy of the modes ($P \parallel {\bf Q} \parallel Z$). Following the results in the Appendix, one can immediately see the effect of the chiral term in each one of the energy shifted triplets: for the right-handed component, $|11 \rangle$, the mode appears at $\hbar\omega_1 = \hbar\omega_0 - g\mu_BH$ and the scattering cross section is  $\sigma_x^{+-} (1 1) \propto M_{1,1} + M_{ch,1}$. For the left-handed component, $|1 \bar{1} \rangle$, the mode appears at $\hbar\omega_{\bar{1}} (1 \bar{1}) = \hbar\omega_0 + g\mu_BH$. The chiral term enters the scattering cross section with the sign reversed which, in the absence of spin anisotropy, leads to a perfect cancellation with the symmetric spin-spin correlation function and yields a null scattering cross-section,  $\sigma_x^{+-} (1,\bar{1}) \propto M_{1,\bar{1}} - M_{ch,\bar{1}} \equiv 0$. By reversing the direction of the magnetic field with respect to \textbf{Q} or by reversing the neutron polarization directions or even by conducting the experiment at neutron energy gain \cite{Note3} the $|1 \bar{1} \rangle$ component can be materialized. The full calculation of the scattering cross sections has been performed by Lovesey \cite{Lovesey84b} with identical results to ours. Figure \ref{0T_H_P}b shows the effect described above for the $\sigma_x^{+-}(1 1)$ scattering channel. As shown in the appendix, if the anisotropy of the spin-spin correlation function is taken into account then $\sigma_x^{+-} (1,\bar{1}) \ne 0$. However for the values of the anisotropy found here in the preceding section, the calculation shows that  $\sigma_x^{+-} (1,\bar{1})$ amounts to 0.7\% of $\sigma_x^{+-} (1,1)$ and therefore should barely show up above background.\\
\begin{figure}[!ht]
%\color{blue}
\begin{center}
\epsfig{file =  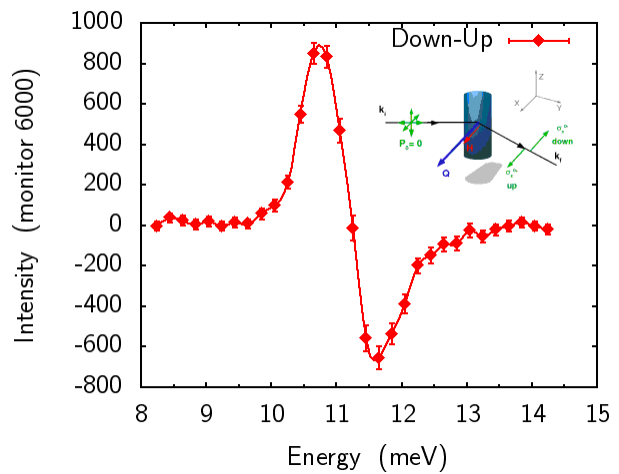, width=8.5cm}
\caption{\label{3T_H_P_0_diff2} (color online) Difference $\sigma_x^{0+} - \sigma_x^{0-}$ for a horizontal magnetic field of 3T (\textbf{Q} $\parallel$ H), following the data in figure \ref{3T_H_P_0}b. Continuous line is a guide-to-the-eye.}
\end{center}
\end{figure}\\
This effect of the neutron beam polarization on the splitted triplet is even more spectacular when the incident beam is unpolarized and the final neutron polarization is analyzed, parallel and antiparallel to the applied magnetic field (and to \textbf{Q}). The corresponding scattering cross sections are now $\sigma_x^{0+}$ and $\sigma_x^{0-}$, respectively. Under these conditions, the $|1 1 \rangle$ component of the triplet appears in the $+$ channel whereas the $|1 \bar{1} \rangle$ appears in the $-$ channel. This is just the consequence of having $-$ and $+$ polarized neutrons along $x$ in the incident beam (neutrons polarized otherwise do not contribute to the measured cross section), respectively. Figure \ref{3T_H_P_0}a shows that for H=0T the scattered intensities are the same for both $+$ and $-$ channels and little can be said on the origin of the signals. When the horizontal magnetic field is applied (Figure \ref{3T_H_P_0}b) each one of the channels displays the effect predicted by the theory. \\

The difference $\sigma_x^{0+} -\sigma_x^{0-}$ (Figure \ref{3T_H_P_0_diff2}) bears an astonishing resemblance to the result proposed as the signature of the \textit{dynamical} chirality in some paramagnetic compounds \cite{Maleyev98,Plakhty99,Plakhty01}, but centered at 11.3 meV instead of at zero energy. Apart from the energy location, our figure is connected to the \textit{trivial} chirality of the magnetic excitations whereas in theirs authors have claimed that it is the signature of chiral magnetic fluctuations induced by a magnetically chiral ground state \cite{Maleyev98,Plakhty99,Plakhty01}, i.e., the \textit{non-trivial} magnetochiral fluctuations. Disclosed from the argument outlined above, the actual contribution to this difference is 
$I_{diff} \equiv M_{1,1} + M_{ch,1} - (M_{1,\bar{1}} + M_{ch,\bar{1}})$, which in our conditions read as $|M_{1,1}| \equiv |M_{ch,1}|$ and identically $|M_{1,\bar{1}}| \equiv |M_{ch,\bar{}1}|$. Therefore I$_{diff}$ is a proper measure of chirality. Note that we have included a different term of chirality for each one of the triplets as, in general, the \textit{non-trivial} chirality may favor one or the other. In the absence of \textit{non-trivial} magneto-chirality both terms are identical, although centered at their respective magnetic field Zeeman splitted positions.

\section{Discussion}

\subsection{Magnetic anisotropies}

Our polarized inelastic neutron scattering experiments have revealed an anisotropy between the magnetic fluctuations along $a^*,c^*$ (or in the plane of the chains) and $b^*$ (the stacking direction). It is important to discuss to which extent magnetic anisotropies observed by other techniques may relate to the evidences here reported. This result, often present in antiferromagnetic Cu-salts, has been found in other Cu$^{2+}$ spin-singlet ground state compounds such as CuGeO$_3$ \cite{Hase93}, BaCuSi$_2$O$_6$ \cite{Sasago97}, and also the chain part of Sr$_{14}$Cu$_{24}$O$_{41}$ \cite{Matsuda96a}. It also appears in compounds where the absence of holes and the near 90$^\circ$ Cu-O bonding leads to the condensation of a ferromagnetic order along the chains and an antiferromagnetic order among the chains. This is the case of Li$_2$CuO$_2$ \cite{Sapina90,Chung03,Boehm98} and La$_{5}$Ca$_{9}$Cu$_{24}$O$_{41}$\cite{Matsuda98} in  which the magnetic moments point along the stacking direction and there is, in addition, a substantial magnetic moment at the oxygen sites, 0.1$\mu_B$ for the former and 0.02$\mu_B$ for the latter.  Experiments under magnetic fields have shown a nearly Ising behavior \cite{Ammerahl00} that underlines the rather strong anisotropy for this spin 1/2 compound. Whether the presence of a magnetic moment on the oxigen is a signature of some circulating currents, the origin of such anisotropy remains unclear yet. 

It is well established that, in the absence of sizeable spin-orbit coupling, the magnetic moment of Cu$^{2+}$ would be isotropic and equal to 1$\mu_B$ ($g=$2). The spin-orbit coupling introduces the mixing of the ground state with the excited states yielding an orbital contribution to the wave function. This orbital contribution $(a)$ modifies the magnetic moment and $(b)$ introduces an anisotropy in the $g-$values, $g_a \ne g_b \ne g_c$. Early ESR experiments on Sr$_{14}$Cu$_{24}$O$_{41}$ \cite{Matsuda96a} have revealed that the Land\'e tensor is anisotropic ($g_a=$2.05, $g_b=$2.26, and $g_c=$2.04) and temperature independent. The values for the anisotropy of the Land\'e factors are typical for a Cu-ion in a square planar coordination of the oxygen ligands. 

A second mechanism that renders anisotropy is the spin-orbit interaction associated to antisymmetric spin-(super)exchange interactions, or Dzyaloshinskii-Moriya interactions. In this case the standart Heisenberg hamiltonian
\begin{equation}
H[S]=J\vec{S}_1 \cdot \vec{S}_2
\end{equation}
transforms into
\begin{equation}
H[S]=(J-\frac{\vec{D}^2}{4J^2})\vec{S}_1 \cdot \vec{S}_2 + \vec{D}\cdot(\vec{S}_1 \times \vec{S}_2) + \frac{1}{2J}(\vec{D}\cdot\vec{S}_1)(\vec{D}\cdot\vec{S}_2)
\end{equation}
with $J=4t^2/U$ the antiferromagnetic superexchange and $\vec{D}=8t\vec{t}/U$ is the Dzyaloshinskii-Moriya vector. $t$ and $\vec{t}$ are the transfer integrals and $|\vec{D}| \propto |\vec{t}| \propto \lambda$, and $\lambda$ the spin-orbit coupling constant. The third term is an anisotropic exchange that is of second order in the spin-orbit coupling. It is this term that provides the anisotropy as $|\vec{D}|/J \propto \Delta g/g$. The caveat of this DM interaction is that it results in a splitting of the triplet of excitations as it is the case in the frustrated spin-dimers SrCu$_2$(BO$_3$)$_2$\cite{Cepas01}. Notwithstanding, Shektmannn, Entin-Wolhman and Aharony (SEA) \cite{Shekhtman92} have realized that there is a hidden symmetry in the hamiltonian above which can be written in the form
\begin{equation}
H[S^{\prime}]=(J+\frac{D^2}{4J})\vec{S^{\prime}}_1 \cdot \vec{S^{\prime}}_2
\end{equation}
with $S_1^{\pm\prime}=e^{\pm i \theta/2}S_1^{\pm}$, $S_1^{z\prime}=S_1^{z}$, $S_2^{\pm\prime}=e^{\mp i \theta/2}S_2^{\pm}$ and $S_2^{z\prime}=S_2^{z}$. This hamiltonian with this definition of the spins is exactly isotropic and therefore has the same eigenvalues and eigenvectors as the previous one. The triplet remains degenerate and anisotropic as required by the experimental results. It was pointed out later that some restrictions apply to SEA's result in that the hidden degeneracy exclusively appears in the case of the one band model and that the degeneracy is raised once the multi-orbital aspect is taken into account \cite{Koshibae93,Shekhtman93}. Moreover, Hund's rule coupling was not considered in the SEA transformation which again will act in the way to raise the degeneracy of the triplet. 

%Early ESR experiments on Sr$_{14}$Cu$_{24}$O$_{41}$ \cite{Matsuda96a} have revealed that the Land\'e tensor is anisotropic ($g_a=$2.05, $g_b=$2.26, and $g_c=$2.04) and temperature independent. The values for the anisotropy are typical for a Cu-ion in a square planar coordination of the oxygen ligands. 

A strong anisotropy of the superexchange in the 90$^\circ$ Cu-O chains has been advanced as a very likely property of this type of chains\cite{Tornow99} which underlines the importance of orbital degrees of freedom. This third possibility has been invoked to explain the very large linewidth of the ESR signal in La$_{14-x}$Ca$_x$Cu$_{24}$O$_{41}$ \cite{Kataev01a,Kataev01b}, and in LiCuVO$_4$ \cite{vonNidda02}, and the inelastic neutron scattering data on the magnetic excitations in Li$_2$CuO$_2$ \cite{Boehm98} and Ca$_2$Y$_2$Cu$_5$O$_{10}$ \cite{Matsuda01,Matsuda05}. ESR studies on pure Sr$_{14}$Cu$_{24}$O$_{41}$\cite{Kataev01b} have disclosed a similar anisotropy as well as a particular temperature dependence of the linewidth, constant up to T$^* \approx$200K and then linearly increasing. This temperature marks the onset of a 1D charge melting and it does not correspond to a real phase transition. Similar drastic changes above T$^*$  have been observed in the spin excitations of the chain sublattice\cite{Regnault99,Matsuda99} and in the temperature evolution of some Bragg reflections\cite{Braden04,Zimmermann04}. However, and contrary to expected, the broadening of the ESR lineshape in Sr$_{14}$Cu$_{24}$O$_{41}$ amounts to 1/100 of that observed in La$_{5}$Ca$_9$Cu$_{24}$O$_{41}$ thus ruling out the influence of
the anisotropic superexchange in Sr$_{14}$Cu$_{24}$O$_{41}$. 

It seems that for this compound the presence of long range ordered Zhang-Rice hole pairs helps stabilizing and strengthening the antiferromagnetic interactions (and from that the presence of a large spin gap), the exchange coupling becomes drastically different as now next-nearest neighbors couplings have to be taken into account as well \cite{Kataev01b}.  Thus the theoretical model casts to describe all the above mentioned compounds\cite{Tornow99} turn out to be inadequate for chain-compounds having a rather strong antiferromagnetic Cu-Cu super-exchange as it is found in pure Sr$_{14}$Cu$_{24}$O$_{41}$. The case of super-exchange coupling mediated by a double-bridge (see figure \ref{Structure_chains}) has been recently considered\cite{Gelle05} following embedded crystal fragment, \textit{ab-initio} cluster calculation. Very importantly, this work shows that the magnetic orbital is essentially supported by the 3d$_{xy}$ Cu orbital with a delocalization tail on the surrounding O 2$p$ orbitals, the average repartition being 2/3 and 1/3, respectively. Such magnetic contribution of the oxygen orbitals is the largest found in spin-chains, at least three times larger than that of Li$_2$CuO$_2$\cite{Chung03}, and has important consequences in the neutron scattering experiments: (a) the magnetic form factor for the spin-spin correlation functions is going to be  very anisotropic due to the planar geometry of orbitals involved\cite{Note4}  and (b) the phase factors of Cu and O will produce interferences that result in rather unusual Q-dependence of the structure factor of excitations.

From the \textit{ab~initio} calculations\cite{Gelle05}, magnetic electrons are broadly distributed within the cluster that results in a large oxygen contribution. This result and the presence of a large anisotropy of the spin-spin correlation functions leads us to the conclusion that the spin-orbit interactions within this cluster are going to be significant, probably more important than in ferromagnetic La$_{5}$Ca$_{9}$Cu$_{24}$O$_{41}$\cite{Matsuda98}.  The occurrence of an orbital moment within the cluster is certainly the signature of uncompensated currents, long predicted for these type of compounds.

\subsection{\textit{Trivial} chirality}

The intrinsic helicity of the spin excitations can, under appropriate conditions of magnetic field, give rise to a non null antisymmetric spin-spin correlation function or dynamical magnetochirality. This feature is considered in the neutron scattering equations and is a general property of any magnetic system. In this paper we have examined the influence of $M_{ch}$ in the scattering cross section of the spin-singlet to spin-triplet transitions in a spin 1/2 dimer compound. When compare with the classical systems, the interest of this quantum spins systems is two fold: (a) Excitations are well defined at sufficiently low temperatures and appear at finite energy. (b) Due to the nature of the excitations, spin singlet-to-triplet transitions can be easily calculated (see Table \ref{tab:table1}) and results are rigorous. Each component of the triplet is going to split following the Zeeman energy term, $\hbar\omega_{1,\bar{1}}=\hbar\omega_0 \mp g\mu_B H$. The components that are important for our purposes are the $|1 \bar{1} \rangle$ and $|1 1 \rangle$, whereas the $|1 0 \rangle$ is dispersionless and lack of further interest and, furthermore, is not visible in the experimental configuration $H \parallel \bf{Q}$. Interestingly, the number of components available for scattering studies in this configuration is reduced to two. The most appealing result of our experiment is that if a polarized neutron beam is created then the chiral term is going to act differently on each one of the components of the triplet depending upon the polarization of the incident beam. Following the results in Table \ref{tab:table1} and in the appendix, the component $|1 1\rangle$ appears in the $\sigma^{+-}$ channel, the \textit{"+"} sign meaning that the arrow of H, of \textbf{Q}, and of $P_0$ point towards the same direction. The other component, $|1 \bar{1}\rangle$, becomes enhanced by $M_{ch}$ in the $\sigma^{-+}$ channel. We want to stress that $M_{ch}$ is entirely part of the scattering cross section of the magnetic excitations, as it is proved by the exact cancellation of the scattering cross-section in the corresponding channels shown in figure \ref{3T_H_P_0} and perfectly reproduced in the equations (see Appendix).

Interestingly, the use of polarized neutrons allows to single out each one of the components of the triplet which thus endorsing detail studies as a a function of pressure, magnetic field, temperature, separately. Indeed in spin-singlet ground state compounds with spin gaps of the order of 10-40 meV, magnetic fields do not allow for a clear separation of each one of the components of the triplet beyond energy resolution. As a result, a large peak is seen in the scattering experiments that can hardly be analyzed. By combining magnetic fields with neutron polarization analysis this difficulty can be easily overcome. A further application of our results is in detecting spin-only molecular crystal field excitation, i.e.  \textit{dimer} (multi-\textit{mer}) physics hidden in the energy spectra of many antiferromagnetic compounds and probably at the origin of specific behavior such as heavy fermion, superconductivity, \textit{etc.}

\subsection{\textit{non-Trivial} chirality}

Throughout this work we have implicitly assumed that the chain sublattice of Sr$_{14}$Cu$_{24}$O$_{41}$ does not support \textit{non-trivial} dynamical chirality. This may not be totally true if the proposal for the presence of ring exchange between in the 90 $^\circ$ Cu-O-Cu bonds is further confirmed\cite{Kataev01a,Kataev01b,Tornow99}. As it is well known from the spin-ladder systems  this ring exchange is produced by a cyclic four-spin exchange and gives rise to a very intricate phase diagram, that includes ground states with vector or scalar chirality. Despite their underlying interest, these phases have not been observed yet. The proposed ring-exchange in the chain sublattice\cite{Kataev01a,Kataev01b,Tornow99} is of different nature, and so is the analogous phase diagram, yet to be determined. These ring exchange may turn out to be enhanced as a result of the strong antiferromagnetic interactions that mediate the Cu-...-Cu super-exchange  making up the dimer in the chain sublattice, as pointed out by Gell\'e and Lepetit \cite{Gelle05}. 

Nevertheless the chiral interaction vector, if any, would be perpendicular to the plane of chains and will thus remain undetectable in our experiments. As it has been worked out by Maleyev\cite{Maleyev02}, the \textit{non-trivial} dynamic magnetochirality scattering cross section arises from the presence of an axial vector interactions and contains the projection of the spin-spin cross product, ${\bf C=S_i \times S_j}$, in the following form:
\begin{eqnarray}
M_{ch}^{C} \propto (\bf{P \cdot C})
\end{eqnarray}
\textit{Non-trivial} dynamical magneto-chiral fluctuations can be seen as \textit{phason}-like (or twist) as well as \textit{amplitudon}-like excitations of the helix (or variations of the pitch of the helix), or soliton type Bloch domain wall\cite{Braun05}. Note that both type of fluctuations of the helix can be understood as equal and unequal variations of the phase between operators ${\bf S_i}$ and ${\bf S_j}$, respectively. Although not much is known about the properties of these modes, we expect that both to be low lying energy modes. In our experiment, and because ${\bf P} \perp {\bf C}$, we do not expect any \textit{non-trivial} dynamic magnetochirality contribution to appear in our neutron scattering experiments, if any. 

A different proposal of \textit{non-trivial} magnetochiral effects arises from \textit{hidden} order parameters that embody electronic degrees of freedom in highly covalent molecules. We have seen in the introduction that a number of proposals have appeared in the literature to explain particular features in high T$_C$ superconductors.  Quantum spin-ladders, the ladder sublattice of Sr$_{14}$Cu$_{24}$O$_{41}$ have been seen to display features that can interpreted as the effect of spin-currents, ring exchange, biquadratic exchange, four-spin exchange, all these terms used in the literature to name the same effect\cite{Honda93,Bremer99}. Besides, the likely anisotropy originating from these electronic degrees of freedom (see the previous paragraph) that translates the occurrence of an orbital moment in a molecular assembly, one can speculate on the advent of \textit{non-trivial} chirality on the singlet (ground state) or the triplet state. Because of the peculiar symmetry properties of these states it can be easily seen that the triplet is even under the exchange of positions 1 and 2 and therefore $M_{ch}^{C} \equiv 0$ for the triplet where the opposite holds for the spin-singlet state. Therefore a magnetic component of chiral origin and perpendicular to the CuO$_2$ ribbons can appear in the elastic channel despite that the total magnetization of the spin-singlet state is null. This has not been observed yet although not too much effort has been put on the detection of this eventuality. Recently, it has been found a magnetic order in the pseudogap phase of high T$_C$ superconductors \cite{Fauque06} that has been interpreted in terms of the circulating currents model proposed by Varma\cite{Varma97}. Whether this very long sought result can be analyzed within the framework of a \textit{non-trivial} chirality in the spin-singlet channel is matter that certainly deserves a closer look.

\section{Conclusions}

In this paper we have carried out a thorough neutron polarization analysis study of the quantum magnetic excitations in the spin-chain compound Sr$_{14}$Cu$_{24}$O$_{41}$. Two main results unfold from our study. First, the spin-spin correlation functions are found to be rather anisotropic whereas the spin triplet remain degenerate within our instrument resolution. Both features are hard to reconcile within the standard, but otherwise simple, picture of magnetic interactions model. We speculate on the origin of this anisotropy as coming from orbital electronic currents that induce an effective orbital moment to the dimer. Finally, we have evidenced the presence of non-null antisymmetric inelastic spin-spin correlation functions under an external magnetic field. The experimental conditions were exactly the same as those set-up for chiral compounds in refs. \cite{Maleyev98,Plakhty99,Plakhty01}. However, the material under consideration is a spin-liquid, paramagnetic compound that exhibits a spin-singlet ground state and the lattice structure supports an inversion center. $P-$symmetry is, therefore, not violated in the ground state and thus this compound is not chiral. The presence of non-null antisymmetric inelastic spin-spin correlation functions is quantitatively accounted for under the basis of singlet-to-triplet molecular crystal field excitations. The fact that \textit{non-trivial} chirality have the same selection rule as \textit{trivial} chirality casts serious doubts on its observation in paramagnetic compounds by polarized inelastic neutron scattering experiments under parallel-to-\textbf{Q} magnetic field.

\begin{acknowledgments}
We wish to acknowledge enligtheneing discussions with J.P. Boucher at the early stages of this work. O. Cepas and T. Ziman have also contributed a lot to the understanding of the origin of the Dzyalozhinskii-Moriya (DM) antisymmetric interactions. The support of the ILL technical staff is greatly acknowledge.
\end{acknowledgments}

\appendix

\section{Appendixes}

Here we calculate the cross sections taking into account eqs. 1 and 2 and Table \ref{tab:table1}. For the case $H \parallel P_0 \parallel \bf{Q}$ the convention goes as follows: $x \parallel Z$, $y \parallel \{X,Y\}\equiv v$ and $z \parallel \{X,Y\} \equiv w$, where $v$ and $w$ represents a linear combination of operators $S_X$ and $S_Y$, with $v$ and $w$ orthogonal. The most obvious choice is $S_v \equiv S_X$ and $S_w \equiv S_Y$. We have dropped the nuclear correlation function, that appears in the non-spin-flip channels, as it is irrelevant in this calculation. In order to account for the anisotropy of the correlation functions two phenomenological parameters, $a$ and $b$, are used in the calculations. 

\begin{eqnarray}
M_y(|11 \rangle) &=& M_y(|1\bar{1} \rangle) \propto a^2(\hbar \omega) \langle S_v \cdot S^\dagger_v \rangle = a^2(\hbar \omega) \nonumber \\
M_z(|11 \rangle) &=& M_z(|1\bar{1} \rangle) \propto b^2(\hbar \omega) \langle S_w \cdot S^\dagger_w \rangle = b^2(\hbar \omega) \nonumber \\
M_{ch}(|11 \rangle) &=& -M_{ch}(|1\bar{1} \rangle) \propto iab(\langle S_v \cdot S^\dagger_w \rangle - \langle S_w \cdot S^\dagger_v \rangle ) \nonumber\\
                    &=& -2a(\hbar \omega)b(\hbar \omega) \nonumber \\
\end{eqnarray}
and $M_{11} \equiv M_y(|11 \rangle) + M_z(|11 \rangle)$, $M_{1\bar{1}} \equiv M_y(|1\bar{1} \rangle) + M_z(|1\bar{1} \rangle)$. As each component of the triplet occurs at different energies (in the presence of a magnetic field) we further define $a_1 \equiv a(\hbar \omega_1)$, $b_1 \equiv b(\hbar \omega_1)$, $a_{\bar{1}} \equiv a(\hbar \omega_{\bar{1}})$, and $b_{\bar{1}} \equiv b(\hbar \omega_{\bar{1}})$. $\hbar \omega_1=\hbar \omega_0 -g\mu_BH$ and $\hbar \omega_{\bar{1}}=\hbar \omega_0 +g\mu_BH$. The scattering cross-sections under polarized neutrons for $|11 \rangle$ and $|1\bar{1} \rangle$ read as follows

\begin{eqnarray}
\sigma_x^{+-}(|11 \rangle) &\propto &  a^2_1 +b^2_1+2a_1b_1 \nonumber \\
\sigma_x^{-+}(|11 \rangle) &\propto &  a^2_1 +b^2_1-2a_1b_1 \nonumber \\
\sigma_x^{+-}(|1\bar{1} \rangle) &\propto &  a^2_{\bar{1}} +b^2_{\bar{1}}-2a_{\bar{1}}b_{\bar{1}} \nonumber \\
\sigma_x^{-+}(|1\bar{1} \rangle) &\propto &  a^2_{\bar{1}} +b^2_{\bar{1}}+2a_{\bar{1}}b_{\bar{1}}  \nonumber \\
\end{eqnarray}
Taking into account the anisotropy $M_z/M_y = b_1^2/a_1^2 =  b_2^2/a_2^2 \approx $1.4, then it is straightforward to calculate the cross sections above

\begin{eqnarray}
\sigma_x^{+-} &\propto &  a_1^2 + 0.007a_{\bar{1}}^2 \nonumber \\
\sigma_x^{-+} &\propto &  0.007a_1^2 + a_{\bar{1}}^2 \nonumber \\
\end{eqnarray}
Note that the contribution of $a_{\bar{1}}$ (compared to that of $a_1$) in the $+-$ polarization channels is rather small (of the order of 0.7\%) and it can be ignored. The opposite applies to $a_1$ in the $-+$ channels. This non-null value of $a_{\bar{1}}$ in the $+-$ channel arises from the anisotropy of spin-spin correlations between the $z$ and the $y$-directions. For many of the purposes we can approximate the above equation by 

\begin{eqnarray}
%              &=& |11 \rangle &+& |1 \bar{1} \rangle \nonumber \\
\sigma_x^{+-} &=& M_{11}+M_{ch} + \overbrace{\left( M_{1\bar{1}}-M_{ch} \right)}^0 \nonumber \\
\sigma_x^{-+} &=& \underbrace{\left(M_{11}-M_{ch} \right)}_0 + M_{1\bar{1}}+M_{ch} \nonumber \\
\end{eqnarray}

\end{document}